\documentclass{elsart}

 \usepackage{graphics}

\usepackage{amssymb}

\begin{document}

\begin{frontmatter}


\begin{center}
\title{Transverse voltages and reciprocity theorem in magnetic fields for high $T_c$ superconductors}


\author{I.\ Jane\v cek$^{1,2}$, P. Va\v sek$^{2,*}$}

\address
{$^1$University of Ostrava, Dvorakova 7, 70103 Ostrava, Czech Republic\\
$^2$Institute of Physics ASCR, Cukrovarnick\'a 10, 162 53 Praha
6, Czech Republic\\}

\end{center}
\vspace*{10cm}
$^*${\it{corresp. author: P. Vasek, Institute of Physics ASCR, Cukrovarnicka 10,\\ 162 53 Prague 6, Czech Republic\\
tel/fax (+4202) 20318586/ (+4202)33343184\\ e-mail vasek@fzu.cz}}

\newpage

\begin{abstract}
We have tested four-point methods of the Hall effect measurement on BiSrCaCuO (2223) polycrystal and also the validity of the magnetic field form of the reciprocity theorem. We found that different types of determination of the Hall resistance using various combination of measured resistances  provide different  value of it.  We have separated  two parts of the resistance combinations, which are even and  odd in magnetic field, respectively.  The odd part, which is equivalent to the Hall effect,  is equal for all  formulae used. The even part of transverse resistance  varies in  different formulae. The magnetic field form of the reciprocity theorem is not valid. Models for explanation of this violation are also discussed. 

\end{abstract}

\end{frontmatter}
\section{Introduction}
Four-point contacts method is usually used for the measurement of electric transport properties. In samples of an arbitrary shape the van der Pauw method \cite{Pauw1}, \cite{Pauw2}, \cite{Pauw3} is often used to determine longitudinal and transversal electric resistivities.
We would like to recall here the main ideas of van der Pauw work. The derivation of basic equations has been made originally for a semiinfinite plane with uniform resistivity $\rho$ in zero  magnetic field ( but similar equations can also be used  in uniform external magnetic field) and with four point contacts (A, B, C, D) placed along boundary. One can get two groups of equations for resistances $R_{kl,mn}$ (in the following the first two indices will mark current contacts, the second pair will denote potential ones) which have the following form:
\begin{equation}
 R{_{kl,mn}} = R{_{mn,kl}}, 	
\end{equation}
\begin{equation}
- R{_{kl,mn}} + R{_{lm,nk}} + R{_{km,ln}} = 0,	
\end{equation}
where k, l, m, n represent arbitrary permutation of the cyclic contacts A, B, C, D. 
Equations  (1) is known also as the reciprocity theorem.  Equations of the second group (2) can be used for calculation of the resistance for selected contact combination from known values of two other resistances obtain for  the other contact combinations. These equations are known from theory of passive multipoles too.\\ 
Results of this derivation can be used for layered samples of arbitrary shape, because semiinfinite plane can be transformed into this shape by conformal transformation. Moreover, it should be emphasized that from basic symmetries of the Ohm's law following equations must be valid:
\begin{equation}
R{_{kl,mn}} = - R{_{lk,mn}} \hspace{1cm}\mathrm {or} \hspace{1cm}  R{_{kl,mn}} = - R{_{kl,nm}}. 	
\end{equation}
Van der Pauw derived also the following relation for resistivity:
\begin{equation}
	\rho = R{_S} t{_S} ,		
\end{equation}
where $ R{_S}$ is the sheet resistance and $t{_S}$ is the thickness of the  sample. The sheet resistance $R{_S}$ can be calculated from two resistances using different cyclic combination of contacts ABCD, for example from $R{_{AB,CD}}$ and $R{_{BC,DA}}$.    $R{_S}$ was defined in \cite{Pauw1}. For isotropic sample  $\rho$  is identical with diagonal elements of resistivity tensor matrix, $\rho = \rho{_{xx}}  = \rho{_{yy}}$ .
If normal Hall effect is present in a sample inserted into magnetic field, we can find  the Hall constant $r{_H}$ from the relation
\begin{equation}
 r{_H}= R{_H}t{_S}/B, 	
\end{equation}
where $B$ is magnetic field perpendicular to the surface of the flat sample and $R{_H}$ is Hall resistance which is  related to the  non-diagonal element of resistivity tensor matrix  $\rho{_{xy}} = R{_H}t{_S}$ .  Principles of the $R{_H}$ determination will be described later.\\
 Eq. (1) represents basic form of the reciprocity theorem: value of the resistivity  $ R{_{kl,mn}}$ does not change after current - voltage contacts interchange.  One can also get   this relation as consequence of the Onsager's relations. But equation in this form  is valid only provided that a  characteristic variable does not change its sign after time reversal. But some variables as magnetic induction $\vec{B}$,  magnetization $\vec{M}$ and others, change sign after time reversion. Thus in magnetic field magnetic field form of the reciprocity theorem                                     
\begin{equation}
 R{_{kl,mn}}(\vec{B} ) =  R{_{mn,kl}}(-\vec{B})   	
\end{equation}
must be valid. In presence of any other additional variable e.g.  $\vec{M}$ (with above mentioned properties)  the general  form of the reciprocity theorem is 
\begin{equation}
 R{_{kl,mn}}(\vec{B}, \vec{M}) =  R{_{mn,kl}}(-\vec{B}, -\vec{M} ) 	
\end{equation}		
On the basis of the microscopic theory the reciprocity theorem was derived by  B\"{u}ttiker \cite{But1}, \cite{But2}, \cite{But3}.\\
Different methods of  measurement of the Hall constant $r_H$ or Hall resistance $R_H$ can be used:
\begin{itemize}
\item[1)] Method using switching-off the magnetic field.
\item[2)] Method of  reversing  the magnetic field.
\item[3)] Method without reversing the  magnetic field based on the validity of reciprocity theorem.
\item[4)] Method without reversing the  magnetic field based on the validity of  equations (2).
\end{itemize}
 Voltage  due to transversal electric field  can be detected using only an arbitrary "cross" combination of contacts (e.g. $R{_{AC,BD}}$ ). The potential contacts in "non-cross" (cyclic) combinations are placed on the same current line at sample boundaries and then they can detect only potential difference   due to longitudinal electric field ( including magnetoresistivity). However,  in a "cross" configuration  one can detect generally not only the transversal field voltage   but  also  longitudinal field voltage  which should be eliminated to get  the transversal field  voltage signal only.  If any contact is not located precisely on sample boundary (or not on the same current line by another reason) the transversal field can be detected also in a cyclic  configuration of the contacts (e.g. $R{_{AB,CD}}$, $ R{_{BC,DA}}$).\\
{\it{Method using switching-off the magnetic field}}. This method is recommended by van der Pauw in his original article \cite{Pauw1}. Hall resistance $R{_H}$ can be obtained from resistance
 $R{_{AC,BD}}(\vec{B})$ in  magnetic field $\vec{B}$
 and from $R{_{AC,BD}}(0)$ measured in zero magnetic field:
\begin{equation}
	R{_H} = R{_{AC,BD}}(\vec{B}) - R{_{AC,BD}}(\vec{0}).		
\end{equation}	
However one should have in mind that this method can be used only when the   magnetoresistance is absent.\\
{\it{Method of reversing the magnetic field}}. In the presence of the magnetoresistance we can use for the elimination of non Hall potentials the fact that the Hall potential contrary to these potentials changes sign (but not absolute value) after reversing the magnetic field. Whence Hall resistance is
\begin{equation}
	R{_H} = [R{_{AC,BD}}(\vec{B}) - R{_{AC,BD}}(-\vec{B})] / 2,  		
\end{equation}	
which is an odd part of cross resistance.\\
{\it{Methods without  change of the direction of the magnetic field}}. From magnetic field form of the reciprocity theorem it follows that $R{_{AC,BD}}(-\vec{B}) = R{_{BD,AC}}(\vec{B})$ and then Hall resistance can be formulated as 
\begin{equation}
	R{_H} = [R{_{AC,BD}}(\vec{B}) - R{_{BD,AC}}(\vec{B}) ] / 2.  	
\end{equation}	
One can also use    Eq. (2)  to detect $R{_H}$ without changing direction of the magnetic field.  Eq. (4)  can be rewritten as:
\begin{equation}
	-R{_{AB,CD}}(0) + R{_{BC,DA}}(0) + R{_{AC,BD}}(0) = 0.		
\end{equation}	
In magnetic field two effects need to be taken into account: the magnetoresistance and Hall effect. In presence of magnetoresistance only  (no Hall effect) this form of the equation could be still valid, because in the van der Pauw derivation we only replace resistivity $\rho(0)$ by resistivity $\rho(B)$ including magnetoresistace. If we consider existence of the Hall effect, we must exclude Hall resistance from "cross" resistance $R{_{AC,BD}} (\vec{B})$. Thus magnetic field equivalent of Eq.(2)  is
$ -R{_{AB,CD}}(\vec{B}) + R{_{BC,DA}}(\vec{B}) + (R{_{AC,BD}}(\vec{B}) - R{_H}) = 0$,  hence
\begin{equation}
	R{_H} = -R{_{AB,CD}} (\vec{B}) + R{_{BC,DA}}(\vec{B}) + R{_{AC,BD}}(\vec{B}).		
\end{equation}	
Similarly, we can obtain also  Eq. (10), because analogically   a magnetic field equivalent of the Eq.(1) can be written as $R{_{AC,BD}}(\vec{B}) - R{_H} = R{_{BD,AC}}(\vec{B}) + R{_H}$. Plus sign  on the right hand side of the equation is due to the fact that the Hall field has opposite direction for the cross resistance $R{_{BD,AC}}$  than for $R{_{AC,BD}}$.
However, if not only the Hall field (transversal electric field that changes sign after reversal of magnetic field), but also field due to "even effect " (transversal electric field which is an even function of $ \vec{B}$)  is present, formulae (9),(10),(12) for the Hall resistance determination do not give the same result.\\ 
The method with reversing magnetic field leads to $R{_H}$, which does not include part due to "even effect". Magnetic field reversal method excludes all even voltages from longitudinal and transversal field. On the contrary, both methods without field reversal described in this section detect all voltages due to transversal electric field.  However, measurement using Eq.(10) can compensate partially or fully both even transversal electric fields added to both "cross" configurations. In this case we must also revert magnetic field to separate Hall effect and "even effect" (detailed analysis of these cases is presented in appendixes A0 - A3).\\
We define following quantities: 
{\it{two-resistance combinations}}
     \begin{equation}
	R^{\mathrm{(II)}}_ {klmn} (\vec{B}) \equiv [R_{kl,mn} (\vec{B}) - R_{ mn,kl} (\vec{B})] / 2
     \end{equation}
and {\it{three-resistance combinations}}
     \begin{equation}
	R^{\mathrm{(III)}}_{klmn} (\vec{B}) \equiv  -R_{kl,mn} (\vec{B}) + R_{lm,nk}(\vec{B})  + R_{km,ln}(\vec{B}),
     \end{equation}
which we use to test different methods of the Hall effect determination and moreover {\it{deviations from magnetic field form of the reciprocity theorem}}
     \begin{equation}
     D_{klmn} (\vec{B}) \equiv [R_{kl,mn} (\vec{B}) - R_{mn,kl} (-\vec{B})] / 2		
     \end{equation}
to test the validity of the reciprocity theorem.  If the theorem is valid $D_{klmn} = 0$.\\ 
We also define odd and even part of resistances:

	$R^{\mathrm{(odd)}}(\vec{B}) = [R (\vec{B}) - R(-\vec{B})]/2$, $\hspace{1cm} R^{\mathrm{(even)}}(\vec{B}) = [R(\vec{B}) + R(-\vec{B})]/2,$
		
where $R$ can represent not only single resistance $R_{kl,mn}$ but also the above defined combinations $R^{\mathrm{(II)}}_{ klmn}$  or $R^{\mathrm{(III)}}_{klmn}$. Odd and even parts obey relation $R^{\mathrm{(odd)}}(\vec{B}) = -R^{\mathrm{(odd)}}(-\vec{B})$  and  $R^{\mathrm{(even)}}(\vec{B}) = R^{\mathrm{(even)}}(-\vec{B})$ , respectively.\\
We analyse the above defined  combinations of  $R{_{kl,mn}}$ measured on cyclic contacts A, B, C, D in the presence of the magnetic field. Details of the analysis are in appendices. Here  we present only the  results.\\
If we assume  the existence of  Hall  and  even effects, we can get following formulae:

$R^{\mathrm{(odd)}}_{AC,BD}  = R{^{\perp,A}}{_{AC,BD}}\equiv R{_H}$

$R^{\mathrm{(II)}}_{ACBD}  =  R{^{\perp,A}}{_{AC,BD}}  +  \triangle R{^{\perp,S}}\equiv R{_H} +$ part due to "even effect",

$R^{\mathrm{(III)}}_{ ABCD} = R{^{\perp,A}}{_{AC,BD}}   + R{^{\perp,S}}{_{AC,BD}}  \equiv  R{_H} +$ part due to "even effect".

Moreover next relations must be valid:

	$R^{\mathrm{(odd)}}_{AC,BD} = R^{\mathrm{(II)~(odd)}}_{ACBD} = R^{\mathrm{(III)~(odd)}}_{ ABCD} = R^{\mathrm{(III)~(odd)}}_{ DABC} = R{_H}$,

	$R^{\mathrm{(II)}}_{ACBD} = [R^{\mathrm{(III)}}_{ABCD} + R^{\mathrm{(III)}}_{DABC}] / 2$.
		
Last relation is fulfil also for odd and even parts, respectively.\\
From the definitions of two-resistances combinations and deviations one gets
\begin{equation}
	R^{\mathrm{(II)~(even)}}_{ACBD} = [D_{ACBD}(\vec{B})  + D_{ACBD}(-\vec{B}) ] / 2\\ 		
\end{equation}
which means that  the parts of the resistance due even transversal electric field are in combination $R^{\mathrm{(II)}}_{ACBD}$ fully compensated in the case of validity of reciprocity theorem.
 The deviations must be equal

$D_{ACBD} (\vec{B})  =  D_{ACBD} (-\vec{B})  =$

\hspace*{1cm}$=[R^{\mathrm{(III)~(even)}}_{ABCD} + R^{\mathrm{(III)~(even)}}_{DABC}] / 2 = R^{\mathrm{(II)~(even)}}_{ACBD}$

which means that, if the reciprocity theorem is valid,  even parts of the three-resistances combination change sign after cyclic permutation. 
\section{Experimental results}
The samples of the BiSrCaCuO, whose zero field data has been  presented in \cite{Java}, was used. The textured polycrystalline sample in form of thin slab has preferred c-axis orientation of the grain perpendicular to the surface.
We have measured three single resistances mentioned in Eq. (14) for basic cyclic configuration of contacts A, B, C, D, and  for cyclic permutation D, A, B, C. In the basic configuration we have  $ R{_{ABCD}}$, 	$R{_{BCDA}}$,  $  R{_{ACBD}}$. For the permutation  we have $ R{_{DABC}}$, $R{_{ABCD}}$ and $ R{_{DBAC}}= - R{_{BDAC}}$.
The  resistances in all configurations were measured in  both directions of the magnetic fields. 
If it is necessary, values for both configurations in one orientation of magnetic field can be measured in one temperature cycle. On the other hand a temperature dependencies for opposite orientation of the magnetic field cannot be simply measured in one temperature cycle.  However, we assume that errors due to small temperature shift discussed in \cite{Java} are negligible (in the same way as in zero field) because the superconducting transition in magnetic field is wider than that in zero magnetic field. Thus for  BiSrCaCuO sample use of data from different measuring run could not be critical.\\
Separated odd and even parts of the  $R^{\mathrm{(II)}}_{ACBD}$, $R^{\mathrm{(III)}}_{ABCD}$  and  $R^{\mathrm{(III)}}_{DABC}$   in magnetic field  $B$ = 1 T are presented on Fig. 1. We can see that odd parts $R^{\mathrm{(II)~(odd)}}_{ ACBD}$, $ R^{\mathrm{(III)~(odd)}}_{ ABCD}$ and $R^{\mathrm{(III)~(odd)}}_{ DABC}$  are  identical. These parts are due to  Hall effect and thus their values equal to Hall resistance $R{_H}$. The temperature dependence of the $R{_H}$  has usual behaviour with  a sign change below the critical temperature. The temperature dependencies of the even parts $R^{\mathrm{(II)~(even)}}_{ACBD}$ and  $R^{\mathrm{(III)~(even)}}_{DABC}$    have similar behaviour with maxima below the  critical  temperature but  the  heights of these maxima are different. The values of the  $R^{\mathrm{(II)~(even)}}_{ACBD}$   approach the theoretical value $( R^{{\mathrm(III)~(even)}}_{ABCD} + R^{\mathrm{(III)~(even)}}_{DACB} ) / 2$.
The temperature dependence of the "cross" resistances $R{_{AC,BD}}$ and $-R{_{DB,AC}}=R{_{BD,AC}} $   for two opposite directions of the magnetic induction are plotted on Fig. 2 and Fig. 3 for magnetic field 1 T and 5 T, respectively. 
The temperature dependencies of  the $R^{\mathrm{(III)}}_{ABCD}$ and $R^{\mathrm{(III)}}_{DABC}$    for several magnetic fields are presented on Fig. 4 and Fig. 5. For low field (up to 0.075 T) $R^{\mathrm{(III)}}_{ABCD}$ and $R^{\mathrm{(III)}}_{DABC}$    are even functions of $\vec{B}$. The odd part is negligible.  For higher fields (1 T and 5 T) even parts and odd parts are plotted separately and are comparable. The odd parts are equal to Hall resistance, which has a well-known temperature and field dependence. Above the critical temperature the Hall resistance is a linear function of B. The even parts have temperature dependence with maxima. Their heights are approximately constant, but the maxima are shifted to lower temperature for increasing magnetic field. The curve for the zero magnetic field has maximum similar to the maximum of the even parts of the $R^{\mathrm{(III)}}_{klmn}$   in  magnetic field. On the other hand this curve has a minimum with negative value similar to minimum, which is observed for small field, in temperature dependence of the Hall resistance (the odd part of the $R^{\mathrm{(III)}}{_{klmn}}$ ). 
Fig. 6 demonstrates breaking of the magnetic field form of the reciprocity theorem in BiSrCaCuO superconductor.  We found the biggest deviation near the critical temperature. However, we can see smaller deviation also above the critical temperature. This deviation decreases with increasing of the temperature. We can see that Eq. (16) is valid. Moreover, the values of the $D{_{ACBD}}(\vec{B})$, $D{_{ACBD}}(-\vec{B})$ are approximately equal to the $R^{\mathrm {(II)~(even)}}_{ACBD}$.
\section{Discussion}
We test different formulae for determination of Hall resistance. Our short analysis of the formulae shows that they may not be equivalent in the presence of "even effect" which was confirmed by the measurement on BiSrCaCuO.  However, after separation of  the even effect we get equivalent values which represent real value of the Hall resistance in agreement with our analysis. Moreover, we find that the reciprocity theorem in the form Eq. (6) is not valid in the BiSrCaCuO polycrystal. This agrees with observed reciprocity theorem breaking in zero magnetic field \cite{Java}. We assume that models suggested there can be extended to the case of the non-zero magnetic field.\\
We suppose that the inequality of the formulae for determination of the Hall resistance as well as the breaking of the reciprocity theorem Eq. (6) is due to the existence of transversal electric field which is even function of magnetic field.
This even field is consequence of an additional anisotropy in sample. In \cite{Java} we suggest two models for  explanation of the  experimental data in zero magnetic field, the model based on guiding of the vortices and the model based on spontaneous magnetization. In case of the BiSrCaCuO  we prefer the former, which changes in presence of the magnetic field to the traditional guided motion model \cite{Kop}, \cite{Sta1}, \cite{Sta2}. This suggestion is supported by behaviour of the even transversal field in magnetic field.\\
In BiSrCaCuO the even effect does not vanish in high magnetic field in contradiction to YBaCuO sample (YBaCuO data will be published later), where reciprocity theorem breaking can be also explained as a consequence of the spontaneous magnetization. Value of this magnetization is too small and thus one can assume that it is not relevant in high magnetic field, but in BiSrCaCuO even effect coexists with  Hall effect ("odd effect").  Existence of small non-zero deviation from the reciprocity theorem above the critical temperature can be due to the existence of superconducting fluctuations.
The transversal effects are usually lower than longitudinal effect in HTSC (and transversal electric fields causes only small deviations in measured resistance (mainly for "cyclic resistance" as e.g. $R{_{AB,CD}}$). However, for ideal cross contact combination  longitudinal field is zero. Our configuration is near this ideal. Longitudinal field is therefore small and deviations due to transversal fields are visible on the curve of the "cross resistance" $R{_{AB,CD}}$.  One can thus see from Fig. 2 and Fig. 3 that reciprocity theorem is not valid:\\
 $R{_{AC,BD}}(+\vec{B},T) \neq R{_{BD,AC}}(-\vec{B},T)$.\\
On the other hand, values of the "cross"  resistance $R{_{AC,BD}}$ (and $R{_{BD,AC}}$, respectively ) in Fig. 2.  after magnetic field reversal  are approximately equal (for field up to 1 T). The difference in resistance for opposite magnetic fields is due to Hall effect , what is clearly visible for higher magnetic fields in Fig. 3.\\ 
We suppose that some general form of the reciprocity theorem should be valid. Such form should incorporate an additional characteristic variable, which changes sign after time reversion.   For the above presented model we suppose in accordance  to models in \cite {Java} internal local  magnetic field due to thermally activated antivortices can play the role of such an additional variable. Thermally activated vortices are captured in lattice of vortices induced by the external  field.   The second model considers spontaneous magnetization as additional variable. The spontaneous magnetization was observed in YBaCuO, but   we have no experimental proof for an existence of spontaneous magnetization in BiSrCaCuO sample.  However, we suppose that the influence of this intrinsic magnetic moment should not be significant for high magnetic field , but we find the reciprocity theorem breaking ( and the presence of even effect as well) in BiSrCaCuO also in high magnetic fields.  

\section{Conclusion}
We show that the different formulae for determination of the Hall resistivity may not give the same value in presence of the even effect. We have tested  these expressions on BiSrCaCuO sample, where we find  that transverse voltage which is an even function of magnetic field (transverse even effect) exists up to 5 T. We assume that observed differences in values of the resistance combinations, used for determination Hall resistance, are due to existence of this even effect. All formulas give the same  value of the Hall resistance after separation of the part of resistances combinations due by the even effect. Moreover we find that magnetic field form of the reciprocity theorem is not valid. This observation is consistent with reciprocity theorem breaking observed in zero-magnetic field.

\section{Acknowledgement}

This work has been supported by GACR under project No.202/00/1602 and by GAAS under project No. A1010919/99 and by Ministry of education under research plan No. 173100003.

\newpage

\appendix{Appendix A0} \\

\begin{tabular}{ll} 
Index $\parallel$ & denotes part of resistance corresponding  to longitudinal electric field $E^{\parallel}$.  \\
Index $\perp$  	 & denotes part of resistance corresponding to  transversal electric field $E^{\perp}$.  \\
Index  S 		& denotes  part of resistance corresponding to symmetric part of the resistivity tensor \\
& (even in magnetic field).  \\
Index A		 & denotes  part of resistance corresponding to antisymmetric part of the resistivity tensor \\
& ( odd in magnetic field).  \\
\end{tabular}

\begin{tabular}{ll}
$R_{kl,mn}$ & 
$= \frac{1}{I}(\int_{n'}^{n}{E^{\parallel}}ds^{\parallel}
+ \int_{m}^{n'}{E^{\perp, S}}ds^{\perp}
+ \int_{m}^{n'}{E^{\perp, A}}ds^{\perp}) = $ \\
& = $ R_{kl,mn}^\parallel + R_{kl,mn}^{\perp, S} + R_{kl,mn}^{\perp, A} $ \\ 
\end{tabular}

The integration path for one combination of contacts is shown on Fig. 7. 

\begin{tabular}{ll}
$R_{kl,mn}^S (\vec{B})  =R_{kl,mn}^{\parallel}(\vec{B}) + R_{kl,mn}^{\perp,S}(\vec{B})$ \hspace{2cm}& 
$R_{kl,mn}^{A} (\vec{B})=~R_{kl,mn}^{\perp, A} (\vec{B})$ \\ 
\end{tabular}

\begin{tabular}{ll}
$R_{kl,mn}^{\parallel} (\vec{B})=~R_{kl,mn}^{\parallel} (-\vec{B})$ 
& $R_{kl,mn}^{\perp,S} (\vec{B})=~R_{kl,mn}^{\perp,S} (-\vec{B}) $ \\ 
$R_{kl,mn}^{S} (\vec{B})=~R_{kl,mn}^{S} (-\vec{B})$ 
&$R_{kl,mn}^{\perp,A} (\vec{B})=-R_{kl,mn}^{\perp,A} (-\vec{B})$  \\
\end{tabular}

\appendix{Appendix A1}

\begin{tabular}{|l|}
\hline 
$R^\mathrm{(even)}_{AC,BD} \equiv [R_{AC,BD}(\vec{B})-R_{AC,BD} (-\vec{B})]/2$ \\
\hline 
\end {tabular}

\begin{tabular}{lll}
$R_{AC,BD}(\vec{B})$ & $= R_{AC,BD}^{S}(\vec{B}) 	+ 	R_{AC,BD} ^{\perp,A}(\vec{B}) $ \\

$R^\mathrm{(odd)}_{AC,BD} $  &	$= [R^S_{AC,BD} (\vec{B}) - R^S_{AC,BD} (-\vec{B})] / 2 +    [R^{\perp,A}_{AC,BD} (\vec{B}) - R^{\perp,A}_{AC,BD}(-\vec{B})] / 2 = $ \\

 & $=   \hspace{2 cm}        0      \hspace{3.3 cm} +     \hspace{1.8 cm}   R^{\perp,A}_{AC,BD} (\vec{B}) $\\
\end{tabular}

\begin{tabular}{l}
$ R^\mathrm{(odd)}_{AC,BD} \equiv R^{\perp,A}_{AC,BD} (\vec{B}) \equiv R_H $ \\
\hline
\end{tabular}\\ \\ \\ \\

\appendix{Appendix A2}

\begin{tabular}{|l|}
\hline
$R^{(\mathrm{II})}_{ACBD} \equiv [R_{AC,BD}(\vec{B})-R_{BD,AC} (\vec{B})]/2$ \\
\hline
\end{tabular}   

\begin{tabular}{ll}
$R_{AC,BD}$ & $ =R_{AC,BD}^{\parallel}(\vec{B}) +  R^{\perp,S}_{AC,BD} (\vec{B}) + R^{\perp,A}_{AC,BD} (\vec{B}) $ \\     
$R_{BD,AC}$ & $ =R_{BD,AC}^{\parallel}(\vec{B}) +  R^{\perp,S}_{BD,AC} (\vec{B}) + R^{\perp,A}_{BD,AC} (\vec{B}) $ \\

$R^{(\mathrm{II})}_{ACBD}$ & $=  [R^{\parallel}_{AC,BD} (\vec{B}) - R^{\parallel}_{BD,AC} (\vec{B})] / 2
+ [R^{\perp,S}_{AC,BD} (\vec{B}) - R^{\perp,S}_{BD,AC} (\vec{B})] / 2 ~+$ \\   
& $+~ [R^{\perp,A}_{AC,BD} (\vec{B}) - R^{\perp,A}_{BD,AC} (\vec{B})] / 2  = $ \\
		
& $= 0   +    \Delta R^{\perp,S} (\vec{B})     +      \Delta R^{\perp,A} (\vec{B}) $ \\

\end{tabular}\\

For isotropic Hall effect   

$R ^{\perp,A}_{BD,AC} (\vec{B}) = R^{\perp,A}_{AC,DB} (\vec{B})  = - R^{\perp,A}_{AC,BD} (\vec{B})$ \\

\underline{$R^{(\mathrm{II})}_{ACBD}  =  R^{\perp,A}_{AC,BD} (\vec{B}) +  \Delta R^{\perp,S} (\vec{B})  \equiv R_H  ~+ $ part due to "even effect"}.\\

For isotropic Hall effect and without "even effect" \\  \\
$R^{(\mathrm{II})}_{ACBD}  =  R^{\perp,A}_{AC,BD} (\vec{B}) \equiv  R_H = R^\mathrm{(odd)}_{AC,BD}$. \\ \\

\appendix{Appendix A3}

\begin{tabular}{|l|}
\hline
$R^{(\mathrm{III})}_{ABCD}  \equiv  -R_{AB,CD} (\vec{B}) + R_{BC,DA}(\vec{B})  + R_{AC,BD}(\vec{B})  $
\hspace{2cm}\\
\hline
\end{tabular}

\begin{tabular}{ll}
$R^{(\mathrm{III})}_{ABCD}$ & $= [-R^{\parallel}_{AB,CD} (\vec{B}) + R^{\parallel}_{BC,DA}(\vec{B})  + R^{\parallel}_{AC,BD}(\vec{B})]  + $ \\

& $+R^{\perp,S}_{AC,BD} (\vec{B}) + R^{\perp,A}_{AC,BD} (\vec{B}) =$ \\
	
& $=     0    ~+$ \\

& $+~ R^{\perp,S}_{AC,BD} (\vec{B}) + R^{\perp,A}_{AC,BD} (\vec{B})$ \\
\end{tabular}

\underline{$R^{(\mathrm{III})}_{ABCD} = R^{\perp,A}_{AC,BD} (\vec{B}) + R^{\perp,S}_{AC,BD} (\vec{B}) = R_H ~+$  part due to "even effect"}.

Without even effect 
$R^{(\mathrm{III})}_{ABCD} 	= R^{\perp,A}_{AC,BD} (\vec{B}) \equiv R_H = R^\mathrm{(odd)}_{AC,BD}$. \\

For  \\ \\
\begin{tabular}{|l|}
\hline
$R^{(\mathrm{III})}_{DABC} \equiv  -R_{DA,BC}(\vec{B}) + R_{AB,CD}(\vec{B})  + R_{DB,AC}(\vec{B}) $\\
\hline
\end{tabular}\\

analogicaly
$R^{(\mathrm{III})}_{DABC} = R^{\perp,A}_{DB,AC} (\vec{B}) + R^{\perp,S}_{DB,AC} (\vec{B})$.

Moreover \\ \\ 
\begin{tabular}{ll}
$R^{(\mathrm{II})}_{ACBD}$ & $= [R^{\perp,S}_{AC,BD} (\vec{B}) - R^{\perp,S}_{BD,AC} (\vec{B})] / 2+
 [R^{\perp,A}_{AC,BD} (\vec{B}) -R^{\perp,A}_{BD,AC} (\vec{B})] / 2=$ \\

& $= [R^{\perp,S}_{AC,BD} (\vec{B}) + R^{\perp,S}_{DB,AC} (\vec{B})] / 2 + 
[R^{\perp,A}_{AC,BD} (\vec{B}) +R^{\perp,A}_{DB,AC} (\vec{B})] / 2 = $ \\

&$= [R^{\perp,A}_{AC,BD} (\vec{B}) + R^{\perp,S}_{AC,BD} (\vec{B}) ] / 2 +
[R^{\perp,A}_{DB,AC} (\vec{B}) +R^{\perp,S}_{DB,AC} (\vec{B})] / 2= $  \\

& $= [R^{(\mathrm{III})}_{ABCD}+R^{(\mathrm{III})}_{DABC}] / 2.$ \\
\end{tabular} \\ \\

\underline{$R^{(\mathrm{II})}_{ACBD} =  [R^{(\mathrm{III})}_{ABCD}+R^{(\mathrm{III})}_{DABC}] / 2$.} \\ \\

\appendix{Appendix B}

\begin{tabular}{ll}
$R^{(\mathrm{II}) ~(odd)}_{ACBD} (\vec{B})$&$  \equiv [R^{(\mathrm{II})}_{ACBD} (\vec{B})  - R^{(\mathrm{II})}_{ACBD} (-\vec{B})  ] / 2=$  	\\
&$= [ R_{AC,BD} (\vec{B}) - R_{BD,AC} (\vec{B}) + R_{AC,BD} (-\vec{B}) - R_{BD,AC} (-\vec{B})] / 4$  \\

$R^{(\mathrm{II}) ~(even)}_{ACBD} (\vec{B})$&$  \equiv [R^{(\mathrm{II})}_{ACBD} (\vec{B})  + R^{(\mathrm{II}) }_{ACBD} (-\vec{B})  ] / 2=$  	\\
&$= [ R_{AC,BD} (\vec{B}) - R_{BD,AC} (\vec{B}) + R_{AC,BD} (-\vec{B}) - R_{BD,AC} (-\vec{B})] / 4$  \\
&$ = \underline{[D_{ACBD}(\vec{B}) + D_{ACBD}(-\vec{B})] / 2}$, \\ \\

\end{tabular}

\begin{tabular}{ll}
$D_{ACBD}(\vec{B})$ & $=~ [R_{AC,BD}(\vec{B}) - R_{BD,AC}(-\vec{B}) ] / 2 =$ \\
& $=~ [R^S_{AC,BD}(\vec{B}) - R^S_{BD,AC}(-\vec{B}) ] / 2 + [R^{\perp,A}_{AC,BD}(\vec{B}) - R^{\perp,A}_{BD,AC}(-\vec{B})] / 2 =$ \\

 & $ =~ [R^S_{AC,BD}(\vec{B}) - R^S_{BD,AC}(\vec{B}) ] / 2 + [R^{\perp,A}_{AC,BD}(-\vec{B}) + R^{\perp,A}_{BD,AC}(\vec{B})] / 2  =$ \\
  & $ =~ [R^{\perp,S}_{AC,BD}(\vec{B}) - R^{\perp,S}_{BD,AC}(\vec{B}) ] / 2 + [R^{\perp,A}_{AC,BD}(-\vec{B}) +
 R^{\perp,A}_{BD,AC}(\vec{B})] / 2$. \\
\end{tabular}

Analogically

$D_{ACBD}(-\vec{B}) = [R^{\perp,S}_{AC,BD}(\vec{B}) - R^{\perp,S}_{BD,AC}(\vec{B}) ] / 2 - [R^{\perp,A}_{AC,BD}(-\vec{B}) + R^{\perp,A}_{BD,AC}(\vec{B})] / 2 $

The second brackets in both expressions are zero for isotropic Hall effect, and then \\ \\
$D_{ACBD}(\vec{B}) = D_{ACBD}(-\vec{B}) = [R^{(\mathrm{III}) ~(even)}_{ABCD}+R^{(\mathrm{III}) ~(even)}_{DABC}]/2=R^{(\mathrm{II}) ~(even)}_{ACBD}$

\newpage
{\large Figure captions}

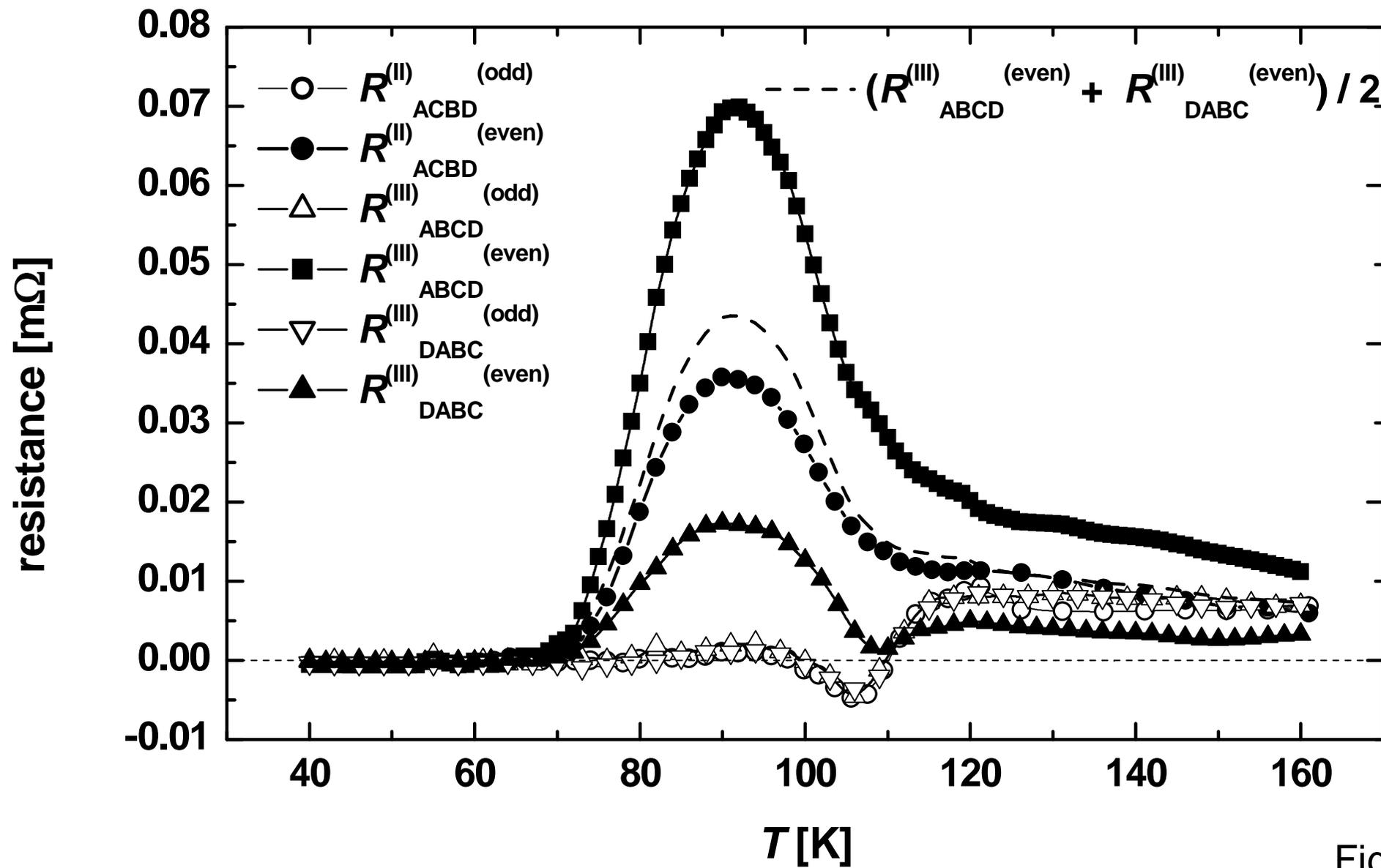
Fig. 1: Temperature dependence of the even and the odd parts of different resistance combinations at magnetic field  1 T.\\
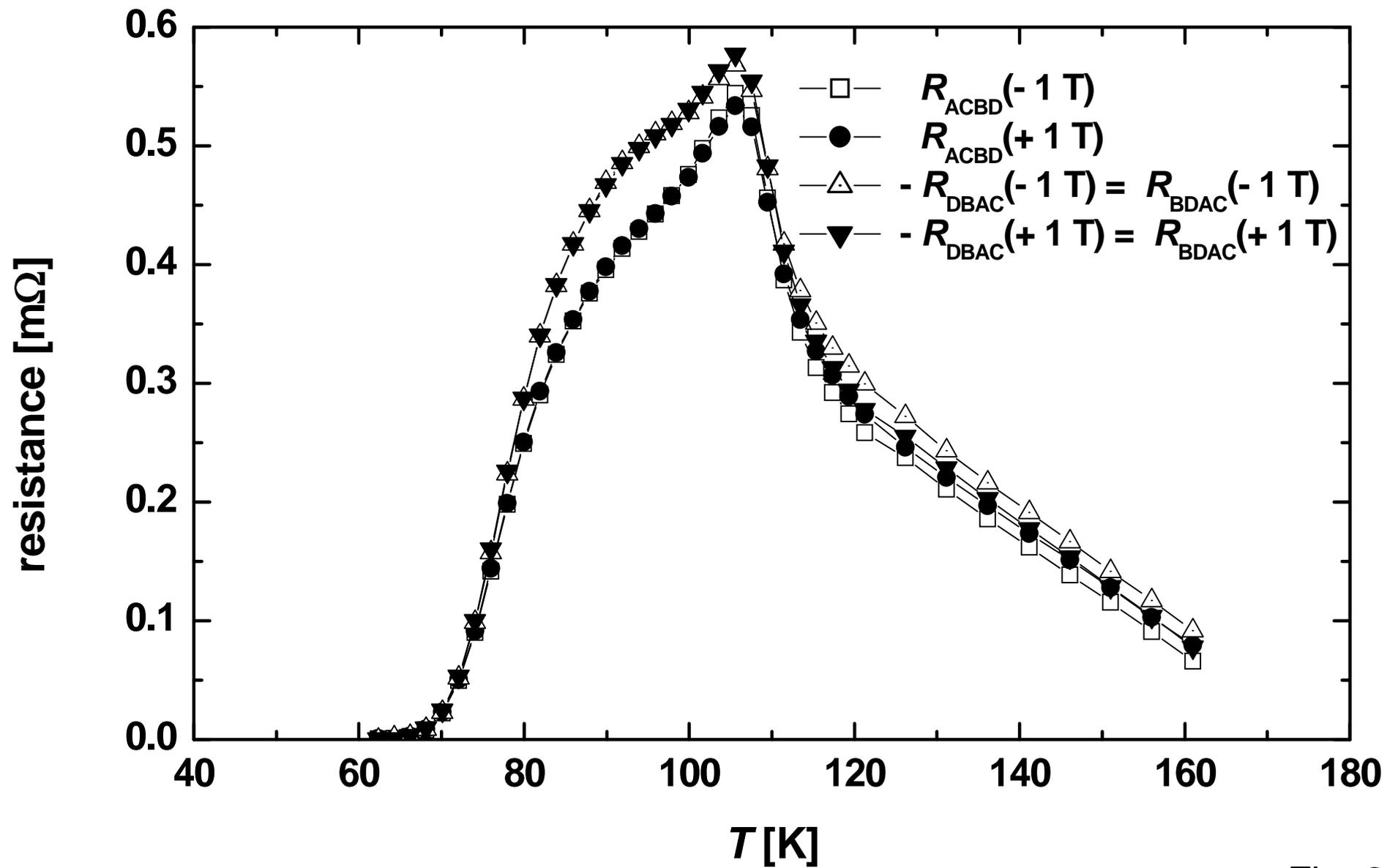
Fig. 2: Temperature dependence of  "cross" resistances in magnetic fields 1T and -1T , respectively.\\
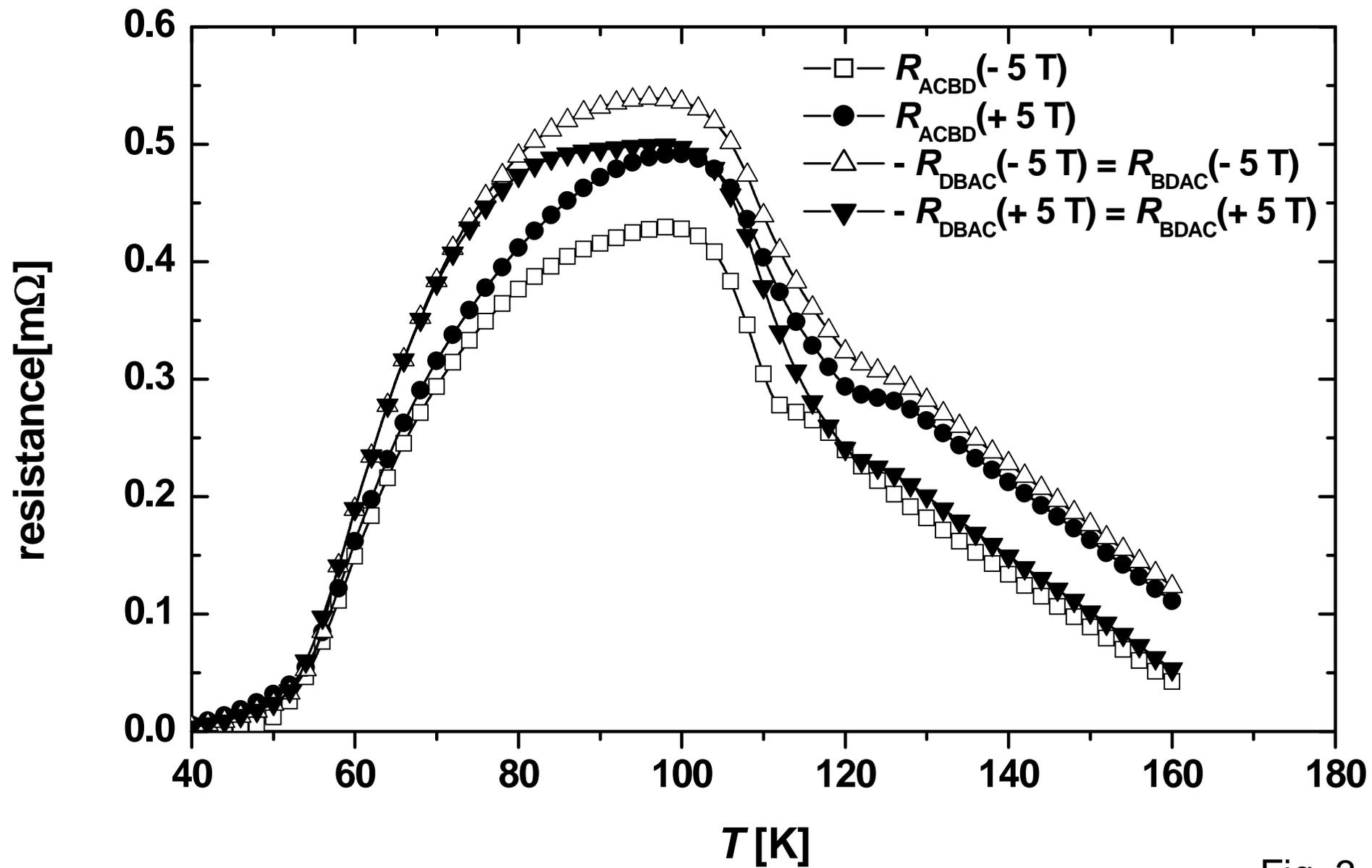
Fig. 3: Temperature dependence of  "cross" resistances in magnetic fields 5T and -5T , respectively.\\
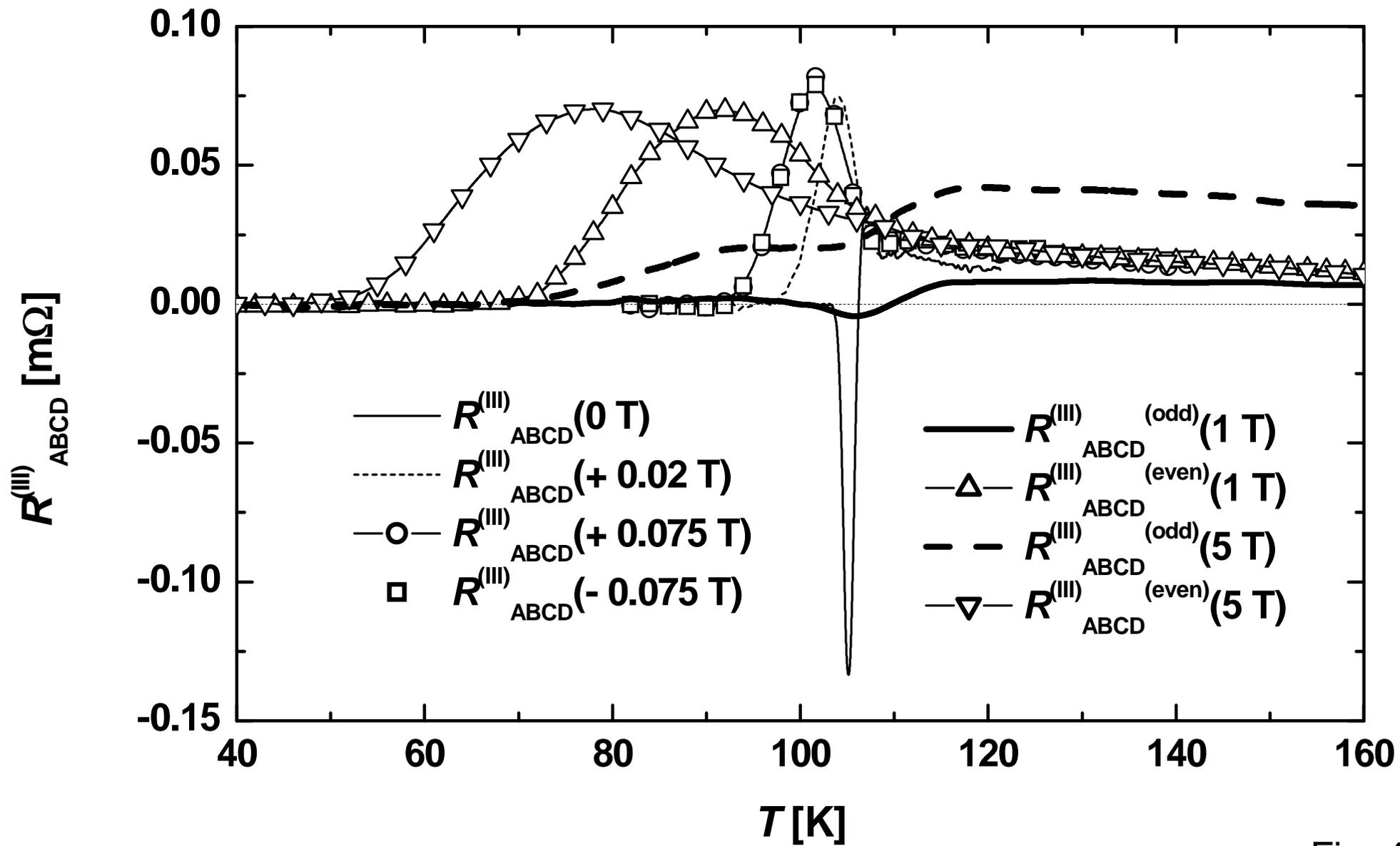
Fig. 4: Temperature dependence of   three-resistance combination $R^{\mathrm{(III)}}_{ABCD}$    for different  magnetic fields .\\
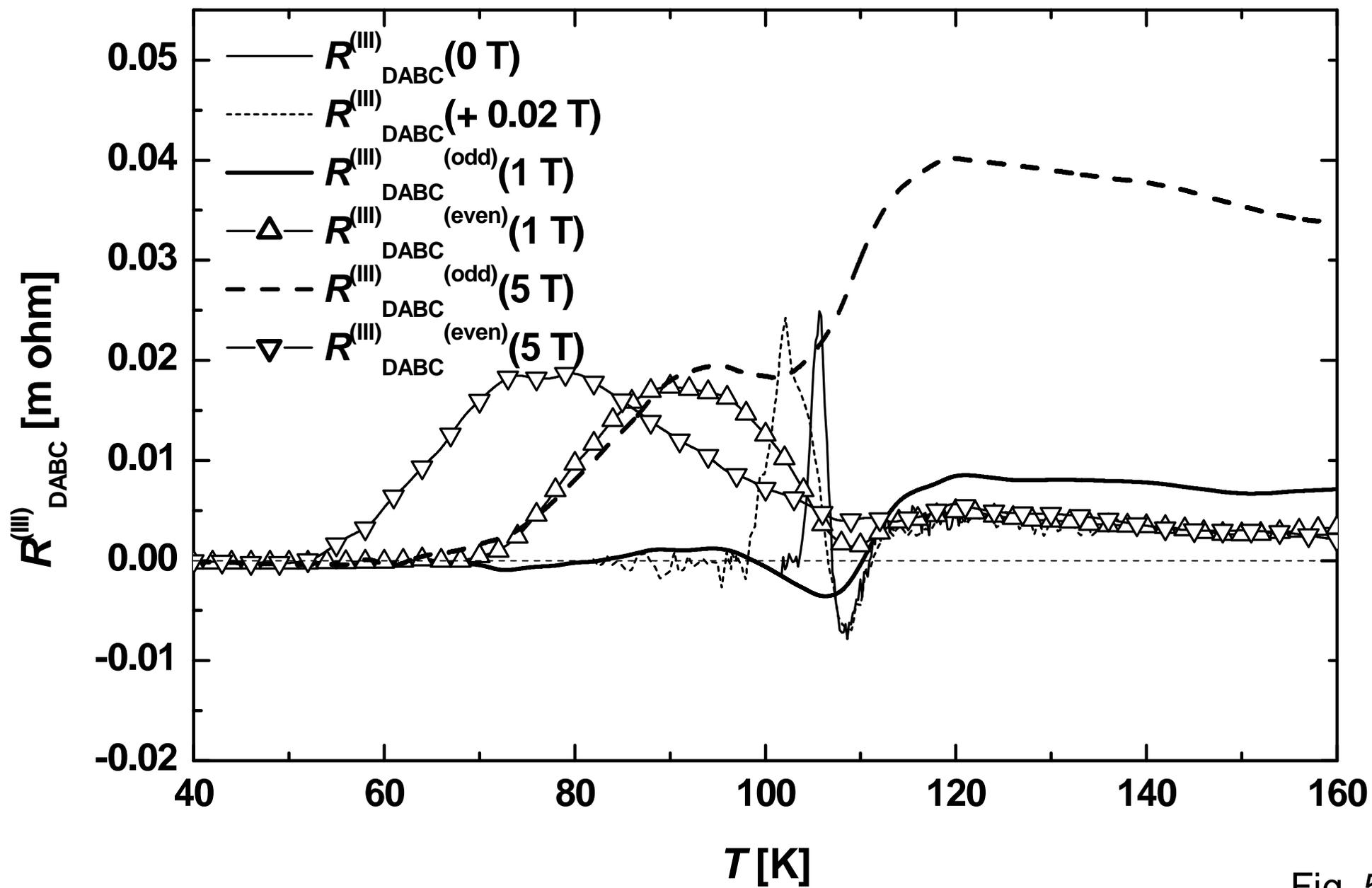
Fig. 5: Temperature dependence of three-resistance combination $R^{\mathrm{(III)}}{_{DABC}}$    for different magnetic fields.\\
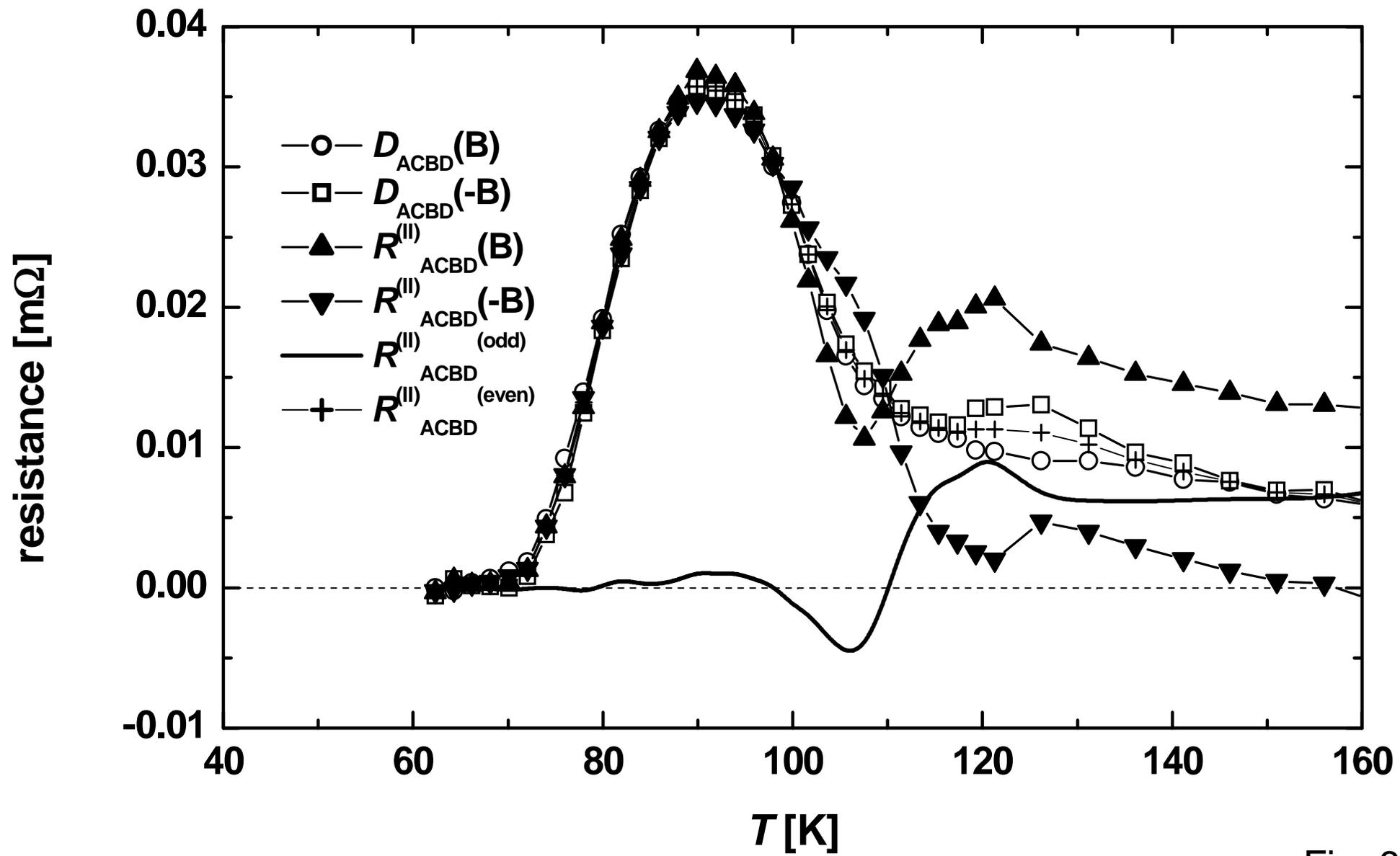
Fig. 6.: Demonstration of the violation of magnetic field form of reciprocity theorem.\\
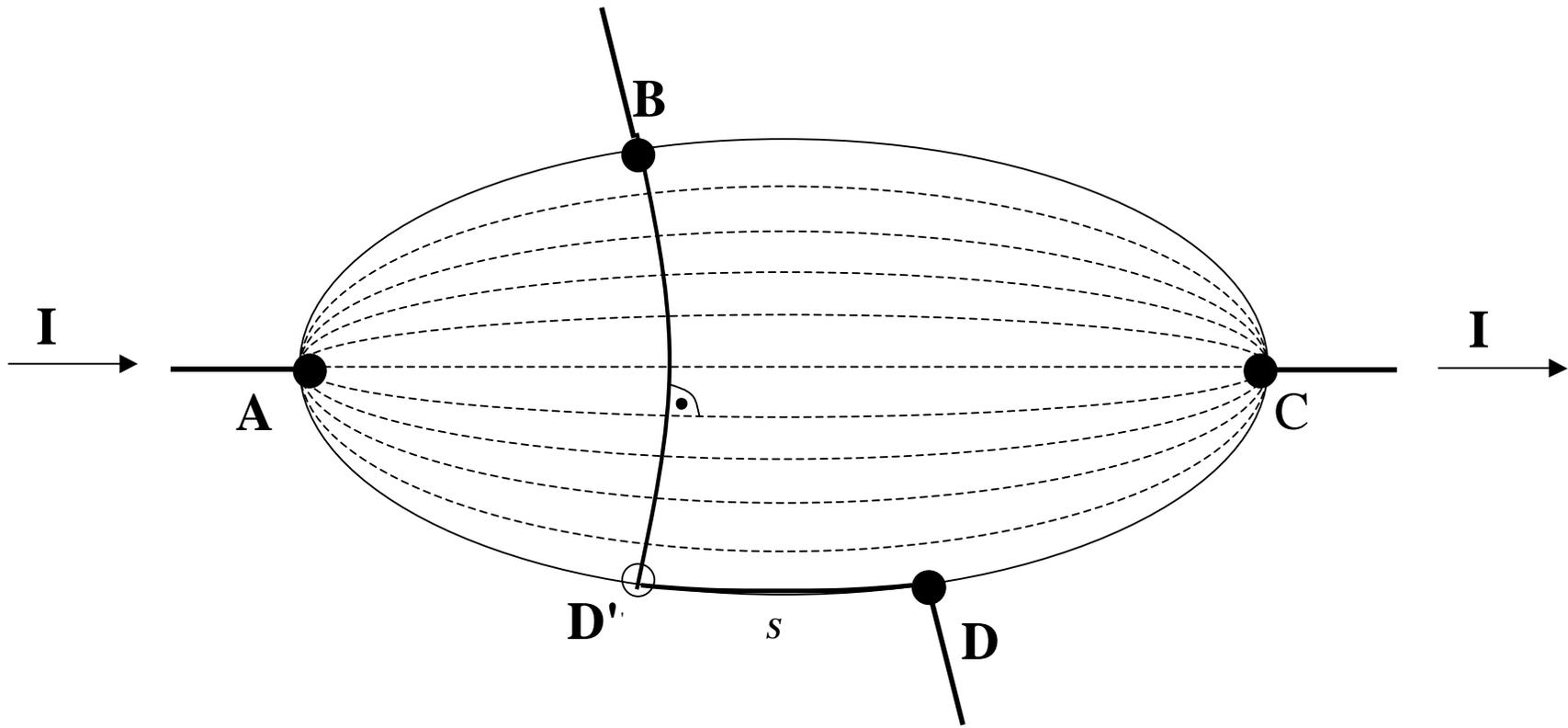
Fig. 7: The integration path used for calculation of  $R_{klmn}$ ( Appendix A0, $kl,mn \equiv AC,BD; n' \equiv D'$) is shown.


\begin{thebibliography}{99}

\bibitem{Pauw1} L. J. van der Pauw, Philips Res. Rep. 13, (1958) 1

\bibitem{Pauw2} L. J. van der Pauw, Philips Tech. Rev. 20, (1959) 220

\bibitem{Pauw3} L. J. van der Pauw, Philips Res. Rep. 16, (1961) 187

\bibitem{But1} M.B\"{u}ttiker, Phys. Rev. B 31, (1985) 6207
	
\bibitem{But2} M.B\"{u}ttiker, Phys. Rev. Lett. 57,  (1986) 1761

\bibitem{But3} M.B\"{u}ttiker, IBM J. Res. Develop. 32, (1988) 317

\bibitem{Java}I. Jane\v{c}ek, P. Va\v{s}ek, Reciprocity theorem in high-temperture superconductors (to be published) 

\bibitem{Kop} J. V. Kopelevitch, V. V. Lemanov, E. B. Sonin, A. L. Kholkin, JETP Letters 50, (1989) 213
 
\bibitem{Sta1} F. A. Staas, A. K. Niessen, W. F. Druyvesteyn, J. van Suchtelen, Phys. Lett., 13, (1964) 293

\bibitem{Sta2} A. K. Niessen, J. van Suchtelen, F. A. Staas, W. F. Druyvesteyn, Philips Res. Repts, 20  (1965) 226


\end{thebibliography}
\end{document}